\documentstyle[aaspp]{article}

\def\stacksymbols #1#2#3#4{\def\theguybelow{#2}
        \def\verticalposition{\lower#3pt}
        \def\spacingwithinsymbol{\baselineskip0pt\lineskip#4pt}
        \mathrel{\mathpalette\intermediary#1}}
\def\intermediary #1#2{\verticalposition\vbox{\spacingwithinsymbol
        \everycr={}\tabskip0pt
        \halign{$\mathsurround0pt#1\hfil##\hfil$\crcr#2\crcr
                \theguybelow\crcr}}}
\def\lta{\stacksymbols{<}{\sim}{2.5}{.2}}
\def\gta{\stacksymbols{>}{\sim}{3}{.5}}

\def\eps@scaling{.95}
\def\epsscale#1{\gdef\eps@scaling{#1}}
\def\plotone#1{\centering \leavevmode
\epsfxsize=\eps@scaling\textwidth \epsfbox{#1}}

\begin{document}
\title{GALACTIC DRIPS
AND HOW TO STOP THEM}

\author{William G. Mathews}

\affil{University of California Observatories/Lick
Observatory\\
Board of Studies in Astronomy and Astrophysics\\
University of California, Santa Cruz, CA 95064\\
mathews@ucolick.org}

\vskip.2in
\begin{abstract}

Under quite general conditions, the temperature of the hot
interstellar gas at large radii in elliptical galaxies can be lower
than the mean galactic virial temperature.  If so, a nonlinear cooling
wave can form in the hot interstellar gas and propagate slowly toward
the galactic core at velocity $\sim 10$ kpc Gyr$^{-1}$.  If the
cooling wave survives hydrodynamic instabilities, it can
intermittently deposit cold gas within about $\sim 15$ effective
radii.  For a bright elliptical the total mass deposited in this
manner is large, $\sim 0.3 \times 10^{10}$ $M_{\odot}$.  In contrast
with {\it ad hoc} assumptions about ``mass dropout'' in galactic
cooling flows, the galactic drip mechanism is a physically self
consistent mechanism in which cold gas can derive from the hot
interstellar gas far from the central regions of elliptical galaxies.
If this cold gas forms into stars, it may account for the young
stellar populations recently observed in many ellipticals.  The
existence of young stars and extended gas at $\sim 10^4$ K observed in
some bright ellipticals may result from this galactic drip rather than
galactic mergers.  Galactic drips are expected in relatively isolated
(field) ellipticals provided (i) the galactic stellar velocity
ellipsoids are radially oriented at large galactic radii and (ii) the
current Type Ia supernova rate is sufficiently small to be consistent
with interstellar iron abundances found in recent X-ray studies.

Ellipticals located within clusters of galaxies are immersed in
generally hotter cluster gas.  If the pressure in the ambient cluster
gas exceeds that in the outer parts of the galactic interstellar
medium, some cluster gas can flow into the galaxy, increasing the gas
temperature there and suppressing the appearence of galactic drips.
Furthermore, the iron abundance in the galactic ISM can be {\it
reduced} when an elliptical is surrounded by hot cluster gas.

\end{abstract}
\keywords{galaxies: evolution, cooling flows, elliptical}
\section{INTRODUCTION}

In this paper we describe a  
global thermal instability and nonlinear cooling wave that can
develop naturally in the hot interstellar medium (ISM) of 
evolving elliptical galaxies.
The ISM in elliptical galaxies
arises from mass loss from an evolving population of old stars.
Gas expelled from mass-losing stars 
into the ISM dissipates the kinetic energy
inherited from the parent star and attains 
a temperature comparable
to the galactic virial temperature 
($T_{\rm vir} \sim 10^{6.5} - 10^7$ K).
As a result the hot ISM occupies approximately
the same volume as the visible stars.
The hot ISM loses energy by radiative losses and slowly 
sinks in the galactic potential; this is a galactic
cooling flow.
Compression of the subsonically 
inflowing ISM by the galactic gravity field
keeps the gas temperature nearly isothermal throughout
most of the flow,
but radiative cooling losses ultimately dominate
at the base of the flow where the plasma density is highest.
The inflowing gas should cool toward a disk if the galaxy is
rotating (Brighenti \& Mathews 1996).

In this paper we describe the evolution of cooling flows
in a typical large elliptical galaxy, demonstrating
that a cooling wave -- or galactic drip -- instability
can be expected quite generally provided (i) the Type Ia supernova
rate is low (as recent X-ray observations suggest)
and (ii) the stellar velocity ellipsoids are significantly
radial at large galactic radii so that the effective stellar
temperature in the outer galaxy is subvirial.
This mass deposition from the ISM 
may have important implications for creating
a subsystem of younger stars (Worthey et al. 1992;
Worthey et al. 1995), in forming
large, coherent filaments of cold gas far from the 
galactic cores 
(Trinchieri \& di Serego Alighieri 1991)
or possibly in creating 
regions of lower X-ray temperature  
(Kim \& Fabbiano 1995).
These galactic features have previously been identified with
the residue of recent galactic mergers,
but galactic drips may provide an alternative explanation.
However, if the elliptical galaxy is located in a relatively 
dense cluster containing high-pressure cluster gas
then the formation of galactic drips is suppressed.

\section{DESCRIPTION OF THE GALACTIC MODEL}

The evolution of galactic cooling flows is described by
the usual equations of gas dynamics with
additional 
source terms (Loewenstein \& Mathews 1987;
Brighenti \& Mathews 1996).
The continuity equation contains a 
source term $\alpha \rho_*(r)$
(gm s$^{-1}$ cm$^{-3}$) representing the rate that new
gas is introduced into the ISM by evolving stars of
local space density $\rho_*$.
Both mass-losing stars and supernovae are sources
of gas, $\alpha = \alpha_* + \alpha_{sn}$,
although $\alpha_* \gg \alpha_{sn}$.
The momentum equation contains a gravitational term,
including an important contribution from dark matter,
and a sink term $- \alpha \rho_* u /\rho$,
where $u$ and $\rho$ are the velocity and density of 
the local ISM.
The sink term accounts for the momentum removed from the 
local flow in accelerating newly injected gas up to
the local flow velocity.

The rate of change of the specific thermal
energy $\varepsilon$ is given by
$$ \rho {d \varepsilon \over dt } =
{P \over \rho} {d \rho \over d t}
- { \rho^2 \Lambda(T) \over m_p^2}
+ \alpha_* \rho_*
\left[ \varepsilon_o - \varepsilon - {P \over \rho}
+ {u^2 \over 2} \right],\eqno(2.1)$$
where $\Lambda (T)$ is the usual cooling coefficient
for an optically thin plasma. 
The term $\varepsilon_o = 3 k T_o / m_p$,
where $m_p$ is the proton mass,
is the specific thermal energy corresponding to 
the mean injection temperature of stellar ejecta:
$$T_o = {\alpha_*(t)T_*(r) + \alpha_{sn}(t) T_{sn}
\over \alpha_* + \alpha_{sn}}.\eqno(2.2)$$
In equation (2.1) $\varepsilon_o - \varepsilon$
represents the excess of total specific kinetic and thermal
energy in stellar ejecta above that in the ambient plasma.
A fraction $P/\rho \varepsilon$ of the energy of newly
supplied gas is lost as the dense stellar ejecta does work
against the ambient gas in achieving pressure balance.
Finally a small amount of energy per gram 
$u^2/2$ is dissipated due to the motion of
the cooling flow gas
relative to the rest frame of the stars.

For simplicity and for comparison with other recent theoretical
studies of galactic cooling flows 
we assume that the stellar mass is described with a
simplified King model, 
$\rho(r) = \rho_{*o}[1 + (r/r_{*c})^2]^{-3/2}$ where 
$r_{*c}$ is the stellar core radius.
The mass distribution in the dark halo is described
with a pseudo-isothermal profile,
$\rho(r) = \rho_{ho} [1 + (r/r_{hc})^2]^{-1}$.
Both density distributions terminate at some large
radius $r = r_t$.
The galactic parameters that we adopt here 
are appropriate for a luminous
elliptical lying on the fundamental plane
(Dressler et al. 1987; Djorgovski \& Davis 1987;
Tsai \& Mathews 1995; Brighenti \& Mathews 1996).
We assume that the total mass in the dark halo $M_{ht}$
is nine times that in the stars $M_{*t}$. 
All relevant galactic parameters are listed in Table 1.

A self-consistent
stellar temperature $T_*(r)$, required to determine
the injection temperature $T_o(r)$,
can be found by solving the equation of stellar 
hydrodynamics in the total galactic potential
(Mathews 1988).
Such a solution requires that some assumption be
made about
the variation of the stellar velocity ellipsoid
with galactic radius.
For this purpose we assume that the stellar orbits
become progressively more radial as $r \rightarrow r_t$.
Specifically, we adopt a velocity ellipsoid of the form
$$ { { \langle v_{r}^2 \rangle - \langle v_{tr}^2 \rangle} \over
{ \langle v_{r}^2 \rangle } }
= \left({ r \over r_t}\right)^{q}\eqno(2.3)$$
where $\langle v_{tr}^2 \rangle $ and 
$\langle v_{r}^2 \rangle $ are the
transverse and radial stellar
velocity dispersions respectively.
We choose $q = 2$,
but the evolution of the ISM is not sensitive to the value of $q$ 
over a wide range $1     \la     q     \la     3$.
A preponderance of radial orbits is expected as a 
general result of many galactic formation models
and is consistent with the anisotropic stellar pressure
required to understand the observed flattening of
massive ellipticals which is not due to galactic rotation.
[We note however that oblate ellipticals can 
also be supported by
increasing the azimuthal stellar velocity dispersion relative
to the mean dispersion in the meridional plane (Satoh 1980).]
However, a global instability in the stellar system 
toward growth of $l = 2$ modes
is expected if the orbits are too radial
(Polyachenko \& Shukhman 1981).
For stability it is necessary that $R_K \equiv 2 K_r / K_t
\la 1.7 \pm 0.2$ where
$K_r = {1 \over 2} \int d^3 v d^3 x v_r~^2f$ 
and $K_t = {1 \over 2} \int d^3 v d^3 x 
(v_{\theta} + v_{\phi})^2f$ are the kinetic energies in the 
radial and transverse directions
(e.g. Bertin \& Stiavelli 1993) 
and $f$ is the stellar distribution function.
For the orbital model described by equation (2.3) with
$q = 2$ we find $R_K = 1.08$, which is safely stable.
Finally,
Tsai and Mathews (1995) have shown that 
mean stellar temperatures
resulting from (2.3) are similar to those recently
measured in the ISM of several bright ellipticals
(Serlemitsos, et al. 1993; Mushotzky et al. 1994).

The rate coefficient for stellar mass loss $\alpha_*(t)$
for an old, single-burst
stellar population can be represented quite accurately
by a power law:
$$\alpha_s(t) = \alpha_*(t_{now})(t/t_{now})^{-p}$$
where $ p = 1.3$ and $\alpha_*(t_{now}) 
= 5.4 \times 10^{-20}$ s$^{-1}$ 
is the current stellar mass loss rate (Mathews 1989)
evaluated at $t_{now} = 15$ Gyr.
Fortunately $ p $ and $\alpha_*(t_{now})$ depend only 
very weakly on the poorly-known IMF (Mathews 1989).

Type Ia supernovae may supply additional energy to the
cooling flow gas.
The equivalent temperature of supernova ejecta in
equation (2.2) is
$T_{sn} = M E_{sn}/ 3 k m_{sn}$ where $E_{sn}$ 
and $m_{sn}$ are the energy and mass
of material expelled in each supernova explosion.
The Type Ia supernova rate observed in elliptical galaxies
$\nu_{sn} = \alpha_{sn} M_{*t}/m_{sn}$ 
is usually characterized in SNu units,
the number of supernovae in a 100 year interval
from a stellar mass having luminosity 
$L_B = 10^{10} L_{B \odot}$.
To represent $\alpha_{sn}$ in cgs units we use the mass to light
ratio determined by van der Marel (1991):
$M_{*t} / L_B = 2.98 \times 10^{-3} L_B    ^{0.35} h^{1.7}$
where $h = H/100$ is the reduced Hubble constant,
taken as $h = 0.75$ here.
Consolidating these results, the factor $\alpha_{sn} T_{sn}$
required for the injection temperature is 
$$\alpha_{sn} T_{sn} = 2.13 \times 10^{-8} 
~( E_{sn} /10^{51} {\rm ergs})~h^{-1.7}     L_B    ^{-0.35}    
~{\rm SNu}(t)~~   {\rm Ks}^{-1}. \eqno(2.4)$$
The factor ${\rm SNu}(t)$ describes the current and
past supernova rate.
The current value ${\rm SNu}(t_{now})$, with $t_{now} = 15$ Gyr,
is poorly known from direct observation
and its past behavior is very uncertain, evidently
depending on the exotic evolution of mass exchanging
binary stars.
Observed values of the supernova rate in elliptical galaxies,
SNu$_{ob}$ $\approx 0.2 h^2$
(van den Bergh \& McClure 1994;
Cappellaro et al. 1993), are also poorly determined and
differ in principle from SNu in equation (2.4) which
is defined relative to other 
uncertain theoretical quantities
(such as $E_{sn}/10^{51}$).
Therefore
SNu$(t_{now}) \sim$ SNu$_{ob}$ is expected, but not 
exact equality.
An upper limit on 
the value of ${\rm SNu}(t)$ in equation (2.4) is provided by 
the iron abundance observed in the hot ISM;
we assume that each Type Ia supernova releases $0.7$ $M_{\odot}$
of iron.
Unfortunately this constraint on SNu is
complicated by the uncertain amount of
additional iron that accompanies normal stellar mass loss
$\alpha_* \rho_*$ into the ISM.
For the past history of Type Ia supernovae we adopt a simple
power law with adjustable parameters:
$${\rm SNu}(t) = {\rm SNu}(t_{now}) (t/t_{now})^{-s}\eqno(2.5)$$ 
with $s = 0.5$ or 1.0 (Loewenstein \& Mathews 1991).
Positive values of $s$ correspond to the decline in supernova
activity expected in cosmic evolution.

Additional iron is supplied to the ISM by normal stellar mass loss.
For this purpose we adopt a stellar iron abundance given by
$$z_*(r) = 1.6 z_{\odot} 
    [1 + (r/r_{c*})]^{-m}  \eqno(2.6)$$
where $z_{\odot} = 1.77 \times 10^{-3}$ is the 
solar abundance.
The iron abundance at the centers of bright ellipticals
is generally thought to be in the range 1 - 2 $z_{\odot}$.
Regarding the exponent $m$,
Thomsen \& Baum (1989) find $z_*(R) \propto R^{-0.3}$
for the magnesium-index (Mg$_2$) gradient in terms of the 
projected radius $R$.
Owing to the rapid decrease of stellar density with
galactic radius,
the variation of metallicity with physical radius,
$z_*(r)$, is almost identical to that 
observed in projected light.
But the iron abundance may not follow that of the
magnesium index which itself is poorly known beyond
an effective radius.
In addition the iron abundances observed in the hot
ISM are often much lower than those in stars within 
an effective radius.
For these reasons we consider 
both $m = 0.3$ and $m = 1.3$ in equation (2.6).

The variation of the iron abundance in the ISM
is described by
$$ {\partial \rho_z \over \partial t}
+ { 1 \over r^2} {\partial \over \partial r}(r^2 \rho_z u)
= (\alpha_* z_* + \alpha_{sn} z_{sn}) \rho_*$$
where $z_{sn} \approx 0.5$ is the fraction of mass in
Type Ia supernova that is iron and
$\alpha_{sn} = 1.52 \times 10^{-17} h^{-1.71} L_B^{-0.354}
{\rm SNu}(t)$ s$^{-1}$.
Throughout the following discussion of the evolution
of the ISM and its iron abundance the only galactic
parameters that are varied are $m$, SNu$(t_{now})$ and
$s$.

\section{THE GALACTIC DRIP PHENOMENON}

The high iron abundance in the intercluster medium 
surrounding massive ellipticals,
$z_{cl} \approx 0.3 - 0.4 z_{\odot}$, indicates that 
strong galactic winds dominated the earliest 
galactic evolution.
Since our interest here is specific to the later evolution
of the ISM when cooling flows dominate, we begin the  
evolutionary calculation after
the initial wind phase is assumed to have ended, at 
$t_1 = 1$ Gyr with a very low initial gas density,
$\sim 10^{-4} \rho_*$.
The spherically symmetric
gas dynamical equations are integrated 
from $t_1$ to the present time, $t = t_{now} \equiv 15$ Gyr
using a time dependent Lagrangian code.
For the calculations described here the boundary condition 
at the outer boundary of the calculation $r = r_t$
is assumed to be a fixed wall,
but calculations done with a free surface at 
$r = r_t$ give essentially the same results in all important
respects.
When gas cools to less than $3 \times 10^4$ K
in a computational zone, it is 
removed from the calculation.

The temperature and 
number density profiles of three cooling
flows are illustrated in Figure 1 at a variety of times for 
parameters 
SNu$(t_{now}) = 0.066$, 0.022, 0.011 all with $s = 1$, i.e.
SNu$(t) \propto t^{-1}$.
For the largest current supernova rate considered, 
SNu$(t_{now}) = 0.066$,
the temperature and density evolve in a perfectly regular
manner (panels a and b of Figure 1), 
just as expected for normal cooling flows.
All the gas cools within the galactic core.
However, when the current supernova rate is reduced to
0.022 (panels c and d), 
the gas temperature in the outer parts of the galaxy 
is smaller until age $t \sim 14$ Gyr 
and profile irregularities appear.
With SNu$(t_{now}) = 0.011$ the irregularities in the flow
propagate inward as seen in Figures 1e and 1f.
In Figure 1e at $t = 6$ Gyrs
a small temperature minimum appears at $\log r \approx 4.85$
which becomes more pronounced at times 8 and 10 Gyrs as it
moves slowly toward the galactic center.
The amplitude of the temperature minimum gradually increases
as the cooling wave propagates inward,
occasionally deepening catastrophically
due to severe radiative losses.
When the gas temperature drops below $3 \times 10^4$ K, 
the gas 
is removed from the hot phase of the ISM and begins to
free fall in the galactic potential as it experiences
a variety of hydrodynamic instabilities.
This is the galactic drip.

Figure 2a illustrates the total mass of cold gas that drops out
of these cooling flows
beyond radius $r$ for the SNu$(t_{now}) = 0.022$ and 0.011
solutions.
In the latter case the mass drop-out 
beyond 5 kpc is large: 
$M_{drop}(r > 5) \approx 2 \times 10^9$ $M_{\odot}$.
This is approximately equal to the total mass of hot gas
remaining in the ISM after 15 Gyr, 
$M_{gas}(t_{now}) = 2.3 \times 10^9$ 
$M_{\odot}$.
Also shown in Figure 2b is are the times and galactic radii
of the specific mass dropouts (where the temperature minimum
drops below $3 \times 10^4$ K) as the galactic drip
moves inward.
For SNu$(t_{now}) = 0.011$ the drip starts at $\approx 72$ kpc
then slowly propagates toward the galactic core 
at velocity $\sim 20$ kpc Gyr$^{-1}$, reaching the center at
$t = 10.5$ Gyr.
None of these details of the cooling flow evolution depend on the
assumed boundary condition at $r = r_t$ (fixed or free),
on the size of the grid zoning,
or on the criterion used to remove cold gas in the drips.

In addition to the deposition of cold gas by the cooling wave,
Figure 2b also indicates that a significant mass of gas cools
within about 5 kpc late in the cooling flow evolution
for both values of SNu$(t_{now})$.
This cooling occurs in numerous small drip events
shown by the large number of points rising vertically at
the left in Figure 2b.
Although located at small galactic radii, this second type of cooling
is well resolved by the numerical gas dynamics.
This cooled 
gas may be optically visible since it is compressed
by the high pressure ISM near the center of the galaxy.

When the past supernova rate is decreased by assuming
SNu$(t) \propto t^{-0.5}$, or $s = 0.5$, galactic drips
are more pronounced for fixed values of the current supernova
rate SNu$(t_{now})$.
Results for these calculations are shown in Figures 3 and 4.
In contrast to the results in Figures 1a and 1b, 
Figures 3a and 3b and 4 show that an appreciable mass of cold
gas drops out within about $r_e =$ 5 kpc from the galactic center
even for the largest current supernova rate SNu$(t_{now}) = 0.066$.
This is evident in the intermittent 
irregularities visible in Figure 3a for times
$t > 10$ Gyr.
The total mass of cold gas that drips out is not small:
$M_{drop}(r > 2 {\rm kpc}) \lta 0.4 \times 10^9$ $M_{\odot}$.
As SNu$(t_{now})$ is decreased to 0.044 and 0.022, 
Figures 3c - 3f and
4 show that the cooling wave begins at earlier times in the 
calculation and at larger galactic radii.
In Figure 3e the cooling wave moves to the origin before the 
time corresponding to the fourth temperature contour
($t = 10$ Gyr),
but the inward motion of the cooling wave 
is nicely resolved in Figure 3c.
In every case after a cooling wave reaches the center of the
galaxy the temperature and density profiles return to 
their normal shapes.

\subsection{Explanation of the Galactic Drip}

Galactic drips arise when gas at great
distances from the galactic center virializes
locally to a temperature that is significantly
less than that of the entire galaxy.
This accumulation of cooler gas at
large galactic radii is possible only if the
supernova rate is sufficiently low.
Most of the gas ejected from stars at large $r$
is deposited early in the galactic evolution
($\alpha_* \propto t^{-1.3}$) when the assumed
Type Ia supernova rate was also larger.
At later times the supernova rate declines and
the temperature of the cool, sub-virial gas
drops further.
Meanwhile the cool gas at large $r$ goes into
approximate freefall and accumulates where it
encounters hot, pressure-supported gas at intermediate
radii.
The density at the base of the approximately freefalling
gas becomes sufficiently large to initiate strong
radiative cooling, a global thermal instability.
As gravity pulls the relatively cool gas toward the
galactic center, adjacent regions just beneath the
cooling gas become compressed
and the non-linear cooling wave slowly propagates
inward in the galaxy, undergoing intermittent ``mass
dropouts'' or ``drips'' when the radiative cooling time
in the wave becomes very short.
In the wake of the cooling wave, the ISM density,
temperature and pressure are all
lowered -- in this sense the cooling wave
resembles a phase change.
Throughout the entire evolution the flow velocity is
subsonic and the gas is in hydrostatic equilibrium to
an excellent approximation.

Just before the cooling wave undergoes one of its episodic
drips, there is a small density inversion in the wave.
Such a condition will result in a localized 
Rayliegh-Taylor instability
which is suppressed in the 1D calculation shown here.
For this reason there is reason to believe that the
evolution of the cooling wave after the first mass drop-out
occurs may be geometrically more complicated than
our 1D results suggest.
If the wavefront becomes distorted, separate regions
of cool gas may appear in the flow.
It is possible, but far from proven, that this could account for 
the numerous cool inclusions which have been recently
resolved by ROSAT PSPC observations in
the outer parts of the cooling flow of the elliptical NGC 507
(Kim \& Fabbiano 1995).
Apart from this possible non-spherical distortion of the flow,
our Lagrangian code is particularly well-suited to compute
the evolution of the thermally cooling wave.
The onset of catastrophic cooling in a small layer of gas
can be masked by the zone-averaging procedures employed in 
Eulerian codes; this may explain why galactic drips have
not been noticed in earlier theoretical calculations.
Nevertheless, Fabrizio Brighenti (private communication)
has successfully duplicated our results using 1D and 2D
Eulerian codes, but only by using a very large number of
computational zones.

The smooth, large scale subsonic velocity field
in cooling flows (not shown in Figs. 1 and 2)
becomes more irregular when the cooling drip occurs.
Low-amplitude sound waves generated near the
nonlinear cooling wave,
particularly during mass dropout episodes,
propagate outward in the galaxy retaining a modest
amplitude as they move
down the galactic density gradient.

\subsection{ISM iron abundance and SNu$(t_{now})$}

The iron abundance in the cooling flow gas depends on
the stellar abundance variation with galactic
radius and the current and past supernova rate.
Recent determinations of the iron abundance in the hot ISM 
in elliptical galaxies are based on the equivalent width
of the complex of FeL iron lines at $\sim 1$ keV.
Abundances observed using the ASCA satellite are very low:
Loewenstein et al. (1994) find $\langle z \rangle
\approx 0.15$ in both NGC 1404 and NGC 4374;
NGC 720, 1399, 4406, 4472 and 4636 have 
$\langle z \rangle \approx 0.3 - 0.4$
(Awaki, H. et al. 1994).
Abundances based on the FeL feature should be accurate
particularly for 
iron ions having only a few bound electrons that are
most abundant at ISM temperatures $T \gta 10^7$ K.
Therefore the FeL abundances should
be reliable for most massive ellipticals.
The mean {\it stellar}
abundance based on equation (2.6) 
is $\langle z_* \rangle = 0.63 z_{\odot}$ if $m = 0.3$
and $\langle z_* \rangle = 0.11 z_{\odot}$ if $m = 1.3$.
These mean stellar abundances provide definite lower limits
on the iron abundance in the ISM since 
the additional iron contributed by supernovae 
can only raise it further.
Obviously, no ISM model computed with 
$z(0) = 1.6 z_{\odot}$ and $m = 0.3$ can match the 
ISM iron abundances measured in NGC 1404 and NGC 4374.
Models
assuming $m = 1.3$ (for which there is no observational
support) do have lower mean iron abundances, 
but still exceed those observed.

Table 2 lists mean iron abundances in the ISM 
$\langle z(r)/z_{\odot} \rangle$ for nine
cooling flow models at time $t = 15$ Gyr.
Very few of these values are as low as those observed 
in the ISM by
the ASCA satellite.
The apparent discrepancy between the stellar 
iron abundances,
the observed Type Ia supernova rate 
and the ASCA iron abundances in the hot gas is an outstanding
problem that is currently unresolved. 
Nevertheless, 
the parameters $s$ and SNu$(t_{now}$ used here to describe
the past and present Type Ia supernova rate
in the cooling flow models
shown in Figures 1 - 4 are not too small;
the discrepancy between the iron abundances found 
in the models shown
in Table 2 and the ASCA values would only be larger
if $s$ and SNu$(t_{now}$ were increased sufficiently
to inhibit galactic drips.
We conclude that
galactic drips should be a common occurrence
in elliptical galaxies.

\subsection{What Happens to the Gas that Cools?}

The final state of the cooled off gas is uncertain,
particularly since magnetic stresses are likely to play
a significant role as the density increases in the
dripping gas.
If we assume nevertheless that the cold gas is able to
form into stars, the total mass of young stars is
small compared to the older stellar population, 
$M_{dt}(r > {\rm 0.5 kpc}) \lta 5 \times 10^{-3} M_{*t}$.
The luminosity of a hypothetical population of young stars
formed from this gas
can be estimated by assuming that they form immediately
after mass dropout and decline in luminosity according
to the single burst luminosity evolution 
(with Salpeter IMF) described
by Bruzual and Charlot (1993).
For most of the ISM calculations with mass dropout
shown in Figures 2 and 4
the resulting approximate
luminosity of young stars is $L_{Bd}(t_{now})
\approx 2 - 6 \times 10^{8}$ $L_{B\odot}$.
This is 
a small fraction of
the total B-band luminosity of the galaxy,
$L_B = 4.95 \times 10^{10}$ $L_{B\odot}$.
In two cases, however, the last mass dropout 
before the calculation was terminated at
$t_{now} = 15$ Gyr is so recent
that the luminosity of young stars is rather large:
$L_{Bd}(t_{now}) \approx 9 \times 10^9$ $L_{B\odot}$
($s = 1$; SNu$(t_{now}) = 0.022$) and
$L_{Bd}(t_{now}) \approx 5 \times 10^9$ $L_{B\odot}$
($s = 0.5$; SNu$(t_{now}) = 0.044$).
Of course these estimates in which 
stars are assumed to form with a normal IMF in this unusual 
environment are very uncertain.
It is possible nevertheless 
that young stars formed in drips could contribute
in an important way 
to the photometric H$\beta$ index 
(a signature of youthful stars) observed 
in many ellipticals by Worthey, Faber and Gonzalez (1992).

The total mass of gas deposited by cooling waves
beyond several kpc, 
$M_{dt} \lta 3 \times 10^9$ $M_{\odot}$,
although small compared to the mass of old stars, is
often comparable with 
the total mass of hot gas remaining in the cooling
flow after $t_{now} = 15$ Gyr, 
$M_{g}(t_{now}) \sim 2 - 6 \times 10^9$ $M_{\odot}$.
If some of this cooled-off
gas resides for a time in pressure equilibrium
with the local hot gas and is kept
ionized by galactic UV radiation (Mathews 1990),
it could account for the large scale,
asymmetric H$\alpha$ filaments extending out to
at least $\sim 10$ kpc observed
in many bright
ellipticals (Trinchieri \& di Serego Alighieri 1991).
The cooling gas in the outer parts of 
galactic cooling flows may also be related to the 
numerous islands of cool gas recently resolved 
by the ROSAT satellite in the
X-ray image of NGC 507 (Kim \& Fabbiano 1995).

\section{ELIMINATION OF DRIPS BY CLUSTER GAS}

For galactic drips to occur it is necessary that
the supernova rates and stellar temperatures be
rather low at large galactic radii,
although both conditions seem easy to justify.
The influence of various supernova rates 
on drip formation is illustrated
in Figures 1 - 4.
To gain insight into the influence of stellar temperature
on galactic drips, several calculations were repeated
with different values of $q$ in equation (2.3).
We find that the cooling 
flow patterns are not critically sensitive to 
this parameter.
In another experiment the galactic stars were divided
into two distributions, one having velocity ellipsoids
as described by equation (2.3) and a second having purely
circular orbits (which have the largest possible temperature
$T_*$ at a given $r$).
Typically we find that galactic drips appear in the flow
unless more than half of the stars 
beyond $\sim 10 r_e$ have highly
tangential (i.e. circular) orbits.
Since geometric flattening in ellipticals depends on a strong
radial component in the stellar orbits, we conclude that
equation (2.3) is a reasonable model.

In addition to adjusting internal parameters, it is also
possible to suppress drips by immersing a drip-prone
galaxy in hot cluster gas.
Even when the velocity of the elliptical is
small relative to the ambient gas,
some hot cluster gas can move into the galaxy,
physically displacing the outer ISM and heating it by 
compression.

To illustrate this latter effect, the 
evolutionary ISM calculation for the elliptical
having parameters $s = 0.5$, SNu$(t_{now}) = 0.022$ and 
$m = 1.3$, which generated drips in Figures 3e, 3f and 4b,
was repeated in a relatively high pressure cluster
environment.
At present we have only a poor understanding
of the evolution of cluster gas, but it is likely
that gas observed in clusters today has evolved 
significantly during the last few Gyr and may
still be highly transient in many clusters.
In order to illustrate the influence of
ambient cluster gas on 
the galactic drip phenomenon,
we assume for simplicity
that the cluster gas has been in existence
with its present-day properties since $t_1 = 1$ Gyr
and that there is no appreciable motion 
of the galaxy relative to the cluster gas.
At $t_1$ the cluster gas is assumed
to be isothermal at $T_{cl} = 10^8$ K and in hydrostatic
equilibrium in the galactic potential before
the evolution of the galactic ISM is begun.
As the galactic ISM evolves, its
cooling flow is born into the
preexisting cluster environment. 
The cluster gas is assumed to extend to 
a fixed wall at 1 Mpc where its
(initial) density is $n_{cl} = 10^{-4}$ cm$^{-3}$.
The initial cluster gas density, in isothermal
hydrostatic equilibrium in the galactic potential,
increases to
$n_{cl} = 2.5 \times
10^{-4}$ cm$^{-3}$ at the galactic center.
These cluster gas parameters have been chosen so that
the radiative 
cooling time beyond the physical limit of the galaxy
($r > r_t$) exceeds 15 Gyr.
During the ensuing evolution the ambient
density and temperature in the cluster gas
($r_t \le r \le 1$ Mpc) remain approximately
constant although some of the cluster gas 
initially beyond $r_t$ flows into the galaxy.
Since the pressure of the ambient hot cluster gas 
does not vary appreciably,
the evolution of the galactic ISM is astronomically 
unrealistic, but we can suppose that the
final state after $t = 15$ Gyr provides a reliable
indication of the present-day influence of the 
cluster environment on the ISM.

Figure 5 compares 
the galactic cooling flow after
$t = 15$ Gyr for the isolated galaxy 
with an identical galaxy that has 
evolved in the cluster environment
as described above.
The increased gas temperature and density
and altered iron abundance
within the galactic boundary (at $\log r_t = 5.05$)
relative to the isolated galaxy (long-short dashed
lines) all arise due to the flow
of cluster gas into the galaxy.
In particular the temperature profile 
in the cluster galaxy 
(solid line in Figure 5a) increases
with radius beyond $r_e$, ultimately joining
the temperature $10^8$ K of the ambient cluster gas.
The gas temperature in the outer parts of the galaxy
is increased due to
thermal mixing with the hot cluster gas that flows into
the galaxy and from the compression of galactic gas 
by the ambient pressure which exceeds that
in the galactic ISM at $r \sim r_t$.
Since the gas temperature is kept high everywhere in 
the flow, no galactic drips occurred in the cluster galaxy 
during its entire evolution 
leading up to the final configuration shown in Figure 5.

When the galaxy is immersed in cluster gas,
the gas density variation 
in the outer galaxy is less steep
(solid line in Figure 5b).
This will result 
in a shallower X-ray surface
brightness profile than that of
the isolated (field) galaxy.

In view of the low iron abundances
in the hot gas determined by ASCA, 
some authors have questioned
whether the metallicity of the 
cluster gas can influence that of the galaxy.
For insight into this question, we illustrate in Figure 5c
the iron abundance at $t = 15$ Gyr in the galactic ISM
for a galaxy having a stellar metallicity 
characterized by a power law index
$m = $1.3 in equation (2.6).
The cluster gas is assumed to have abundance
$z/z_{\odot} = 0.4$.
It is clear from Figure 5c
that the overall ISM metallicity of cluster
galaxies {\it can} be influenced by the flow of
cluster gas into the galaxy.
The iron abundance 
$z(r)/z_{\odot}$
for the isolated galaxy
(long-short dashed line) is actually {\it higher} than that
for the same galaxy
when placed in the iron-rich cluster environment (solid line)!
The explanation of the sense of this difference
is that the inward flow velocity is increased somewhat
when the galaxy is surrounded by high pressure cluster gas.
The iron abundance in the faster moving 
cooling flow gas is lower
since it experiences fewer supernova events during
its shorter flow time to the galactic center.
Paradoxically, 
therefore the iron abundance in an elliptical ISM can be
{\it reduced} by placing it in a hot cluster
of even higher iron abundance!

\section{DISCUSSION AND SUMMARY}

Galactic drips can occur 
in isolated ellipticals whenever (i) the Type Ia supernova
rate is low enough to be (more) consistent with observed iron 
abundances in the ISM and (ii) provided the
stellar orbits become preferentially radial
in the outer galaxy,
leading to relatively low
stellar temperatures $T_*(r)$.
This latter condition is consistent with the asymmetric 
stellar pressure required to flatten ellipticals
having very low rotation velocities.
When these conditions are satisfied, 
a global thermal instability can form in the ISM,
initiating a cooling wave that propagates
slowly inward.
Should finite aspherical density or temperature perturbations 
with appreciable amplitudes exist
in the wave region, the subsequent evolution of the wave 
may be geometrically complex, 
but it is not clear why such perturbations
would be expected.
If the cooling wave develops in a quiesent
galactic atmosphere, as we assume here,
it slowly propagates toward the galactic center, experiencing episodes
of precipitous cooling (drips) which 
can deposit substantial masses 
($\sim 0.2 - 0.3 \times 10^{10}$ $M_{\odot}$) of
cold ($T \la 10^4$K) gas within about 15 $r_e$.
In addition to this sequence of drips, 
a significant amount of gas can cool 
within the brightest parts of the elliptical, $r \lta r_e$,
in a multitude of drips confined to this region.
If the cooled-off gas eventually forms into stars, as many authors
have assumed, it may be sufficient to
account for the populations of young stars reported
by Worthey, Faber, \& Gonzalez (1992).
Unfortunately,
the physics of such a star formation process remains obscure.

In many theoretical models of spherical galactic cooling flows
it has been assumed that cold gas can condense directly from the hot
ISM in thermally unstable regions
located far from the galactic center
(Stewart et al. 1984; Thomas et al. 1987; White \& Sarazin 1987a,
1987b, 1988; Bertin \& Toniazzo 1995).
This assumption was found to be
useful to reduce the theoretical X-ray surface brightness
near the galactic cores to agree
with surface brightness
distributions observed in three bright Virgo ellipticals
(Canizares, Fabbiano, \& Trinchieri 1987).
Unfortunately it has subsequently been shown that low amplitude
perturbations are not generally thermally unstable (Balbus 1991)
while finite amplitude perturbations disrupt rapidly
as a result of shear and Rayleigh-Taylor instabilities
(Hattori \& Habe 1990; Malagoli et al. 1990).
In addition,
there is no natural way for finite
perturbations to arise in the ISM (Mathews 1990).
It is more likely that the 
effects of galactic rotation (Brighenti \& Mathews 1996)
and magnetic field, which are not included in the studies above,
must be considered in comparing theoretical and observed
surface brightness distributions near galactic cores.
However, the galactic drip phenomenon described here does provide
a physically plausible mechanism for thermal instabilities
to develop directly from the hot gas at large galactic radii.
Nevertheless drips cannot provide a general explanation for
the relatively flat central X-ray surface
brightness distributions for ellipticals
since the cooling wave that produces the drip
is a transient effect that cannot be synchronized in all galaxies.
Moreover, it seems likely that the
environmental influences of companion galaxies and
randomly infalling gas in the (poorly understood)
outermost parts of real ellipticals could introduce a
non-spherical character to the initial growth of galactic drips.
In addition, perfectly spherical ellipticals are rare in nature.
Therefore in the absence of 
multidimensional theoretical studies,
it can be doubted that cooling flows in real ellipticals
evolve in the idealized spherical manner shown in Figures 1-4
all the way to the galactic centers.
Since it is likely that cooling waves may not be spatially
coherent on spherical surfaces,
the evolution of the
soft X-ray luminosity and surface brightness profiles
corresponding to Figures 1-4 has not been emphasized here.

Galactic drips can be suppressed in
three ways: (i) by increasing the current or 
past theoretical supernova rate,
(ii) by assuming that a large fraction (at least half) of 
the galactic stars have tangential orbits at large radii
($r \gta 10 r_e$),
or (iii) by immersing the galaxy in hot 
cluster gas.
The first means of avoiding drips is unacceptable
since the global iron abundance would be too high at
the current time, even if the stellar iron abundance drops
much faster with galactic radius than is commonly thought.
The second means of suppressing drips, by introducing
tangential stellar orbits at great distances, may be 
inconsistent with 
models of elliptical formation that have a significant 
dissipationless character.
But in any case, strongly radial orbits are required
to account for the anisotropic stellar pressure support 
that prevails in luminous, non-spherical ellipticals.
Therefore the primary mechanism to avoid galactic drips
may be cluster membership.
If star formation results from galactic drips, field ellipticals
should exhibit more evidence of recent star formation 
(larger H$\beta$ indices) than cluster members.
It is therefore quite remarkable that precisely this
result has been reported by 
Worthey, Trager, \& Faber (1995).
These authors find that the incidence of a population
of younger stars is almost (perhaps totally) a property
of field ellipticals
while cluster ellipticals seem to have little or no
evidence of youthful stars.

This last result may be understood alternatively 
as a consequence of 
galactic mergers which are expected to be more common
for field galaxies.
Although very clear evidence for recent mergers 
exists in some ellipticals,
it is far from proven that mergers are the dominant
ongoing evolutionary process in these galaxies at the
present time or in the recent past.
We have shown here that the appearance of young stars,
transient filaments of ionized gas and cool X-ray inclusions 
are not unambiguous
signatures of recent mergers, but may instead result from
galactic drips, a natural development in the 
evolution of galactic cooling flows.
Finally, we note that the iron abundance in galactic
cooling flows can be {\it lowered} when the galaxy
is located in high pressure, iron-rich  cluster gas; this may
account in part for some of the remarkably low iron
abundances that have been observed in the hot ISM of 
some ellipticals.

WGM thankfully acknowledges support from Calspace, a Faculty
Research Grant from UCSC and NASA grant NAG 5-3060.
Thoughtful remarks and advice from Fabrizio Brighenti,
Sandra Faber, Mike Loewenstein,
Scott Trager, and John Tsai are acknowledged with thanks.

\noindent
\centerline{\bf REFERENCES}\\
\noindent
Awaki, H. et al. 1994, PASJ, 46, L65\\
Balbus, S. A. 1991, ApJ, 372, 25\\
Bertin, G., \& Stiavelli, M. 1993, Rep. Prog. Phys., 56, 493\\
Bertin, G., \& Toniazzo, T. 1995, ApJ, 451, 111\\
Brighenti, F. \& Mathews, W. G. 1996, ApJ (in press)\\
Bruzual, A. G., \& Charlot, S. 1993, ApJ, 405, 538\\
Canizares, C. R., Fabbiano. G. \& Trichieri, G. 1987, ApJ, 312, 503\\
Cappellaro, E. et al. 1993, A\&A, 273, 383\\
Djorgovski, S., \& Davis, M., 1987, ApJ, 313, 59\\
Donnelly, R. H., Faber, S. M.. \& O'Connell, R. M. 1990, ApJ, 354, 52\\
Dressler, A., Lynden--Bell, D., Burstein, D., Davies, R.L., Faber, S.
M., Terlevich, R. J., \& Wegner, G. 1987, ApJ, 313, 42\\
Hattori, M. \& Habe, A. 1990, MNRAS, 242, 399\\
Kim, Dong-Woo, \& Fabbiano, G. 1995, ApJ, 441, 182\\
Loewenstein, M., \& Mathews, W. G. 1987, ApJ, 319, 614\\
Loewenstein, M. \& Mathews, W. G., 1991, ApJ, 373, 445\\
Loewenstein, M., Mushotzky, R. F., Tamura, T., Ikebe, Y., et al. 1994,
ApJ, 436, L75\\
Malagoli, A., Rosner, R. \& Fryxell, B. 1990, MNRAS, 247, 367\\
Mathews, W. G. 1988, AJ, 95, 1047\\
Mathews, W. G. 1989, AJ, 97, 42\\
Mathews, W. G. 1990, ApJ, 354, 468\\
Mushotzky, R. F., Loewenstein, M., Awaki, H., Makishima, K.,
Polyachenko, V. L., \& Shukhman, I G. 1981, Sov. Astron.-AJ,
25, 533\\
Satoh, C. 1980, PASJ, 32, 41\\
Serlemitsos, P. J., Loewenstein, M., Mushotzky, R. F., Marshall, F. E.
1993, ApJ, 413, 518\\
Stewart, G. C., Canizares, C. R., Fabian, A. C., \& Nulsen, P. E. J.
1984, ApJ, 278, 536\\
Thomas, P. A., Fabian, A. C., \& Nulsen, P. E. J. 1987, NMRAS, 228, 973\\
Thomsen, B., \& Baum, W. A. 1989, ApJ, 347, 214 \\
Trinchieri, G., \& di Serego Alighieri, S. 1991, AJ, 101, 1647\\
Tsai, J. C., \& Mathews, W. G. 1995, ApJ, 448, 84\\
van den Bergh, S. \& McClure, R. D. 1994, ApJ, 425, 205\\
van der Marel, R. P., 1991, MNRAS, 253, 710\\
White, R. E. \& Sarazin, C. L. 1987a, ApJ, 318, 612\\
White, R. E. \& Sarazin, C. L. 1987b, ApJ, 318, 621\\
White, R. E. \& Sarazin, C. L. 1988, ApJ, 335, 688\\
Worthey, G., Faber, S. M., \& Gonzalez J. J., 1992, ApJ, 398, 69\\
Worthey, G., Trager, S. C., \& Faber S. M. 1995 (in preparation)\\

\newpage
\begin{figure}
\epsscale{0.80}
\centerline{\plotone{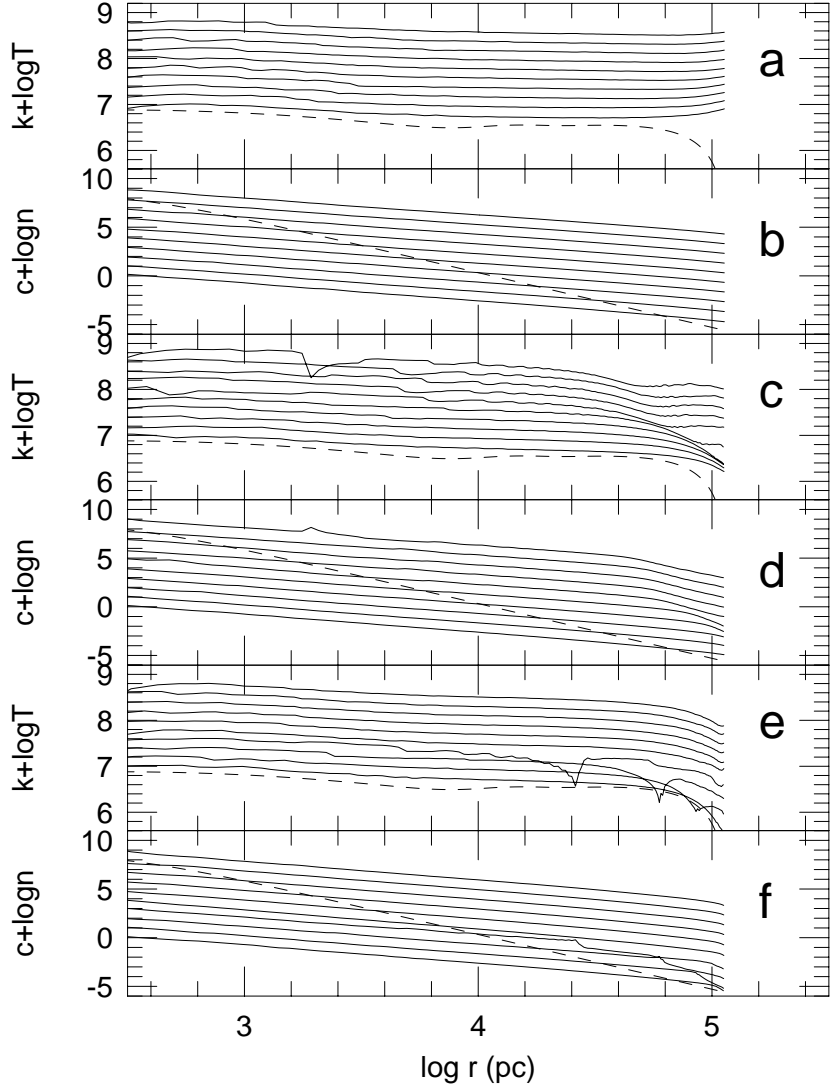}}
\caption{
Temperature and number density evolution in
the ISM of a massive
elliptical when $s = 1$ shown for three values of SNu$(t_{now})$.
Profiles of $T(r,t)$ and $n(r,t)$ are shown at ten times:
4, 6, 8, 10, 12, 13.91, 14.31, 14.51, 14.61, and 15.0
Gyrs.
For clarity
the profiles have been shifted vertically;
the constants on the vertical
axes are $k = 0.2(n-1)$ and $c = 0.1(n-1)$
where $n = 1 - 10$ corresponds to
each of the ten times above.
The stellar temperature $T_*(r)$ (normalized with $k = 0$)
and density $\rho_*(r)$ (unormalized) are shown with dashed lines.
{\it (a and b):} Temperature and density variation with SNu$(t_{now})
=
0.066$;
{\it (c and d):} Temperature and density variation with SNu$(t_{now})
=
0.022$;
{\it (e and f):} Temperature and density variation with SNu$(t_{now})
=
0.011$.
}
\end{figure}

\newpage
\begin{figure}
\epsscale{0.80}
\centerline{\plotone{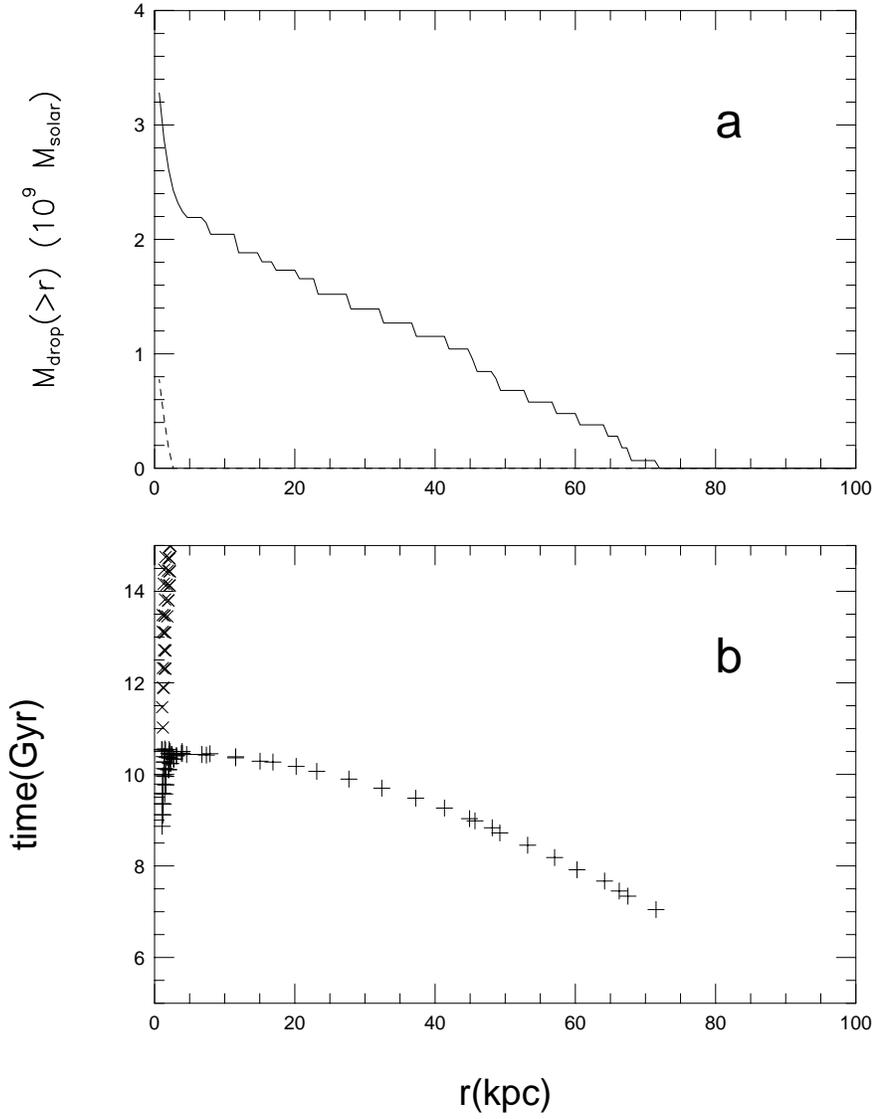}}
\caption{
ISM evolution with $s = 1$. 
(a) Total mass of gas dropped out beyond radius $r$,
$M_{drop}(>r)$ when SNu$(t_{now}) = 0.022$ (dashed line)
and SNu$(t_{now}) = 0.011$ (solid line).
(b) Radii and times of individual drip events in $r > 0.5$ kpc
when SNu$(t_{now}) = 0.022$ (crosses)
and SNu$(t_{now}) = 0.011$ (plus signs).
}
\end{figure}

\newpage
\begin{figure}
\epsscale{0.80}
\centerline{\plotone{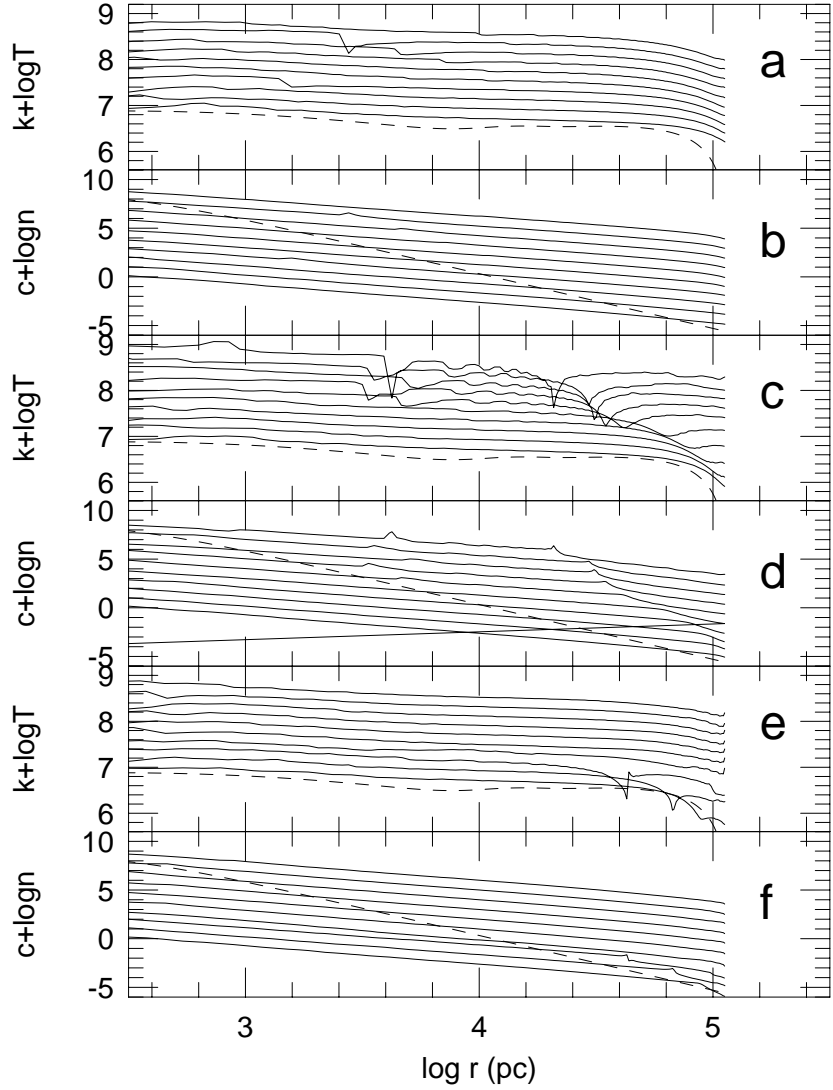}}
\caption{
Temperature and number density evolution in 
the ISM of a massive
elliptical when $s = 0.5$. 
All times and stellar profiles are identical to those in Figure 1.
{\it (a and b):} Temperature and density variation with SNu$(t_{now}) =
0.066$;
{\it (c and d):} Temperature and density variation with SNu$(t_{now}) =
0.044$;
{\it (e and f):} Temperature and density variation with SNu$(t_{now}) =
0.022$.
}
\end{figure}

\newpage
\begin{figure}
\epsscale{0.80}
\centerline{\plotone{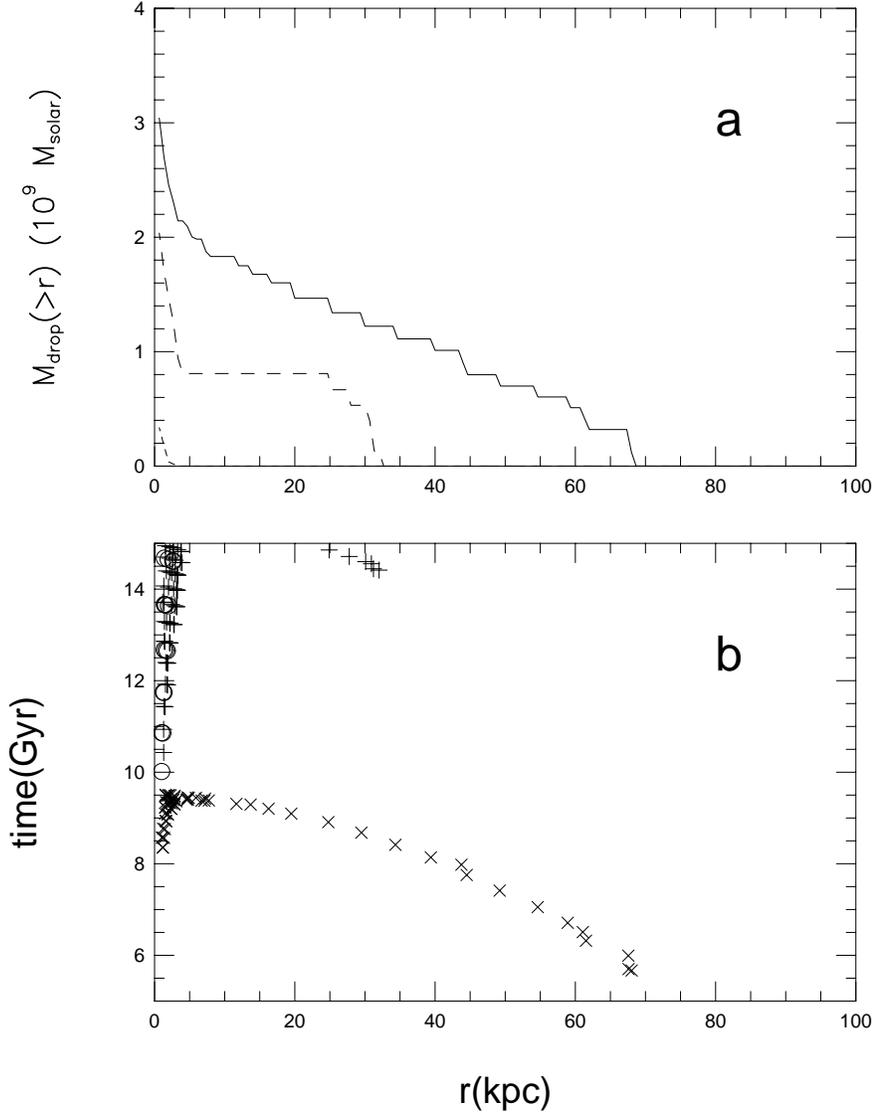}}
\caption{
ISM evolution with $s = 0.5$.
(a) Total mass of gas dropped out beyond radius $r$,
$M_{drop}(>r)$ when SNu$(t_{now}) = 0.066$ (short dashed line)
SNu$(t_{now}) = 0.044$ (long dashed line)
and SNu$(t_{now}) = 0.022$ (solid line).
(b) Radii and times of individual drip events in $r > 0.5$ kpc
when SNu$(t_{now}) = 0.022$ (crosses), 
SNu$(t_{now}) = 0.044$ (plus signs)
and SNu$(t_{now}) = 0.066$ (open circles).
}
\end{figure}

\newpage
\begin{figure}
\epsscale{0.80}
\centerline{\plotone{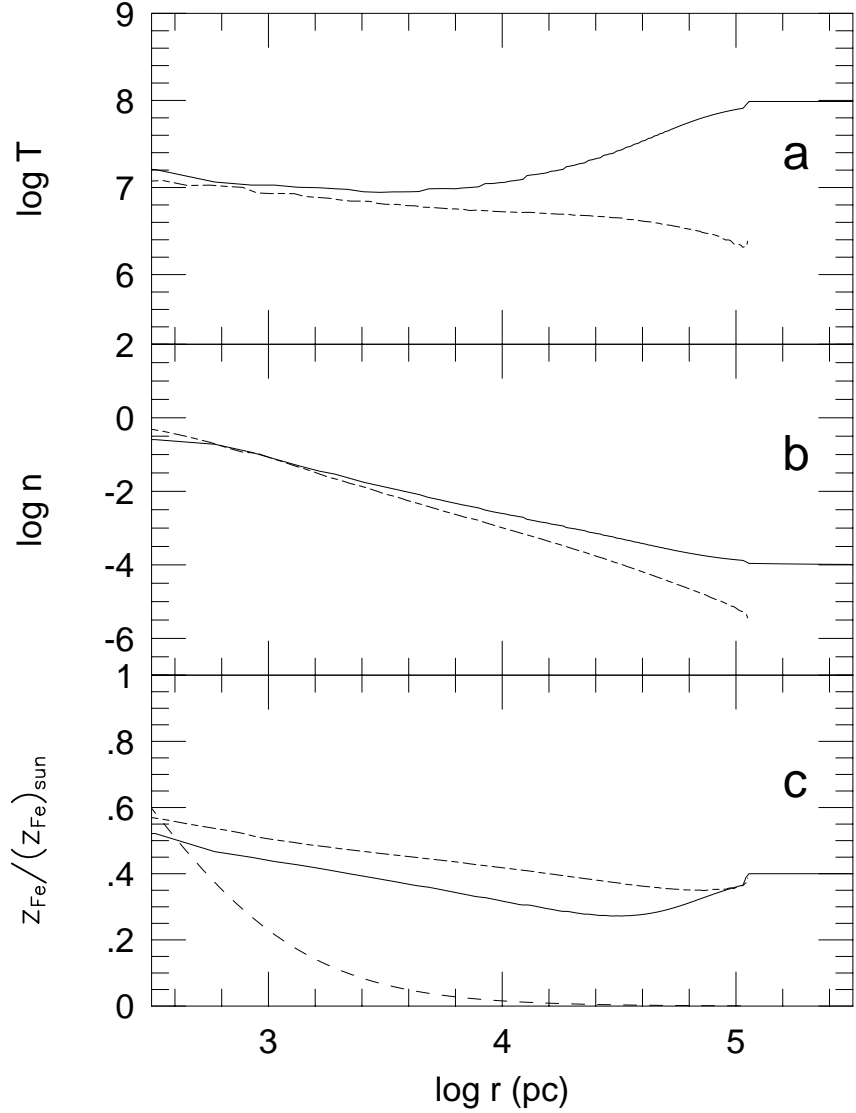}}
\caption{
Influence of environmental cluster gas
on the current properties of a galactic cooling flow.
The galactic parameters are
$s = 0.5$, SNu$(t_{now}) =$ 0.022, and $m = 1.3$.
The cluster gas has temperature $T_{cl} = 10^8$ K
and density $n = 10^{-4}$ cm$^{-1}$ at 1 Mpc.
In all plots solid lines refer to a galaxy immersed
in cluster gas and the long-short dashed lines refer
to the same galaxy when isolated.
(a) Variation of gas temperature with radius,
(b) Variation of gas density with radius,
(c) Variation of the ISM iron abundance (in solar units) 
with radius, including that of the stars $z_*(r)$ 
(dashed lines).
}
\end{figure}

\end{document}